\documentclass[aps,prb,twocolumn,showpacs,twoside,10pt]{revtex4-1}
\usepackage[utf8]{inputenc}
\usepackage{amssymb,amsmath}
\usepackage{graphicx}
\usepackage{color}
\usepackage[colorlinks,bookmarks=false,citecolor=blue,linkcolor=blue]{hyperref}
\usepackage{bbm} 
\usepackage{subfigure}

\newcommand{\et}{{\it et al. }}
\newcommand{\bea}{\begin{eqnarray}}
\newcommand{\eea}{\end{eqnarray}}
\newcommand{\be}{\begin{equation}}
\newcommand{\ee}{\end{equation}}
\newcommand{\nn}{\nonumber \\}

\newcommand{\si}{\mathbf{S_{i\phantom{j}}}}
\newcommand{\sj}{\mathbf{S_{j\phantom{i}}}}

\newcommand{\miintz}{\iint\!\! dz d\bar{z} \,}

\newcommand{\p}{\partial}

\newcommand{\bz}{\bar{z}}
\newcommand{\pbz}{\partial_{\bar{z}}}

\newcommand{\bw}{\bar{w}}
\newcommand{\pbw}{\partial_{\bar{w}}}

\newcommand{\identity}{\mathbbm{1}}

\begin{document}
\title{Analysis of the phase transition for the Ising model on the frustrated square lattice}
\author{Ansgar Kalz}
\email{kalz@theorie.physik.uni-goettingen.de}
\affiliation{Institut für Theoretische Physik, Universität Göttingen, 37077 Göttingen, Germany}
\author{Marion Moliner}
\email{marion.moliner@kit.edu}
\affiliation{Institut für Nanotechnologie, Karlsruher Institut für Technologie, 76344 Eggenstein-Leopoldshafen, Germany}
\author{Andreas Honecker}
\affiliation{Institut für Theoretische Physik, Universität Göttingen, 37077 Göttingen, Germany}
\date{Received 1 June 2011; revised manuscript received 17 August 2011; published 9 November 2011}

\begin{abstract}
We analyze the phase transition of the frustrated $J_1$-$J_2$ Ising model with antiferromagnetic nearest- and strong next-nearest neighbor  interactions on the square lattice. Using extensive Monte Carlo simulations we show that the nature of the phase transition for $1/2 <  J_2/J_1 \lesssim 1$ is not of the weakly universal type -- as commonly believed -- but we conclude from the clearly doubly peaked structure of the energy histograms that the transition is of weak first order. Motivated by these results, we analyze the phase transitions via field-theoretic methods; i.e., we calculate the central charge of the underlying field theory via transfer-matrix techniques and present, furthermore, a field-theoretic discussion on the phase-transition behavior of the model. Starting from the conformally invariant fixed point of two decoupled critical Ising models ($J_1 = 0$), we calculate the effect of the nearest neighbor coupling term perturbatively using operator product expansions. As an effective action we obtain the Ashkin-Teller model.
\end{abstract}


\maketitle

\section{Introduction}
The \emph{simple} Ising model which adds up the interactions of two-state variables on a $D$-dimensional lattice has served as a pioneer in many physical problems, especially in statistical mechanics and solid-state physics.\cite{P:ising25} It was one of the first models to mimic the magnetic exchange interactions in condensed-matter theory, and the behavior of phase transitions was studied extensively for this model in different dimensions and by numerous approaches. Nevertheless there are still open questions concerning this model, in particular in two and three dimensions. In two dimensions the model with nearest-neighbor (NN) interactions undergoes an ordering process at a finite temperature which is well understood and establishes the Ising universality class for second-order phase transitions.\cite{B:baxter82} However, for frustrating interactions the phase diagram of the model becomes richer and the physics of the occurring phase transitions becomes more complicated. 

In the present work we focus in particular on the phase transition from the high-temperature paramagnetic phase into an antiferromagnetic collinear phase that is favored by strong additional interactions on next-nearest-neighbor (NNN) bonds, i.e., $J_2 > J_1/2$. This transition also attracted a lot of interest in the past. In the late 1970s first renormalization-group calculations and Monte Carlo (MC) simulations for the Ising model with frustrating interactions were performed by Nightingale\cite{P:nightingale77} and Swendsen and Krinsky\cite{P:swendsen79} and later on by Oitmaa\cite{P:oitmaa81} and Landau and Binder.\cite{P:lanbin80, P:landau80,P:lanbin85,B:lanbin00} They assumed a continuous phase transition and computed for this particular model transition temperatures and critical exponents. Throughout the 1980s it was commonly accepted that the exponents are weakly universal but vary for different degrees of frustration; i.e., only the scaling relations for the exponents are fulfilled but the absolute values are not universal. However, a continuous phase transition with non-universal exponents is only possible if the central charge of the underlying conformal field theory (CFT) is $c\geq 1$.\cite{B:Friedan1984} Meanwhile theories with discrete $c<1$ define universality classes with universal exponents such as the two-dimensional Ising model.\cite{Belavin_1984, Belavin_1984_2} Since the present system can be described by two copies of Ising models in one part of the phase diagram, the universality of the phase transition was under debate. In 1993 López \et presented a mean-field calculation for the model \cite{P:lopez93, P:lopez94} where they find a first-order transition for a finite parameter region of $0.5 < J_2/J_1 \lesssim 1.1$. Recent MC simulations by the group of Malakis \et contradict this scenario at least for the value $J_2/J_1  = 1$.\cite{P:malakis06} On the other hand, our MC results in Refs.~\onlinecite{P:kalz08, P:kalz09} strengthen the scenario of a first-order transition for small values of $0.5 <  J_2/J_1 \leq 0.7$. 

Here, we will show that the first-order scenario is valid up to $J_2 \leq 0.9~ J_1$. However the MC simulations do not give a conclusive picture for larger values of $J_2$ due to increasing length scales. Hence, it was necessary to apply further techniques to understand the nature of the phase transition for these parameters. To compute the central charge of the model we performed a finite-size analysis of the free energy which was calculated via transfer matrix techniques, but we can only get reasonable results for large $J_2/J_1 > 1$. In a last step we start from the limit of two decoupled Ising models (see Fig.~\ref{f:unit_a} below) with antiferromagnetic coupling $J_2$ and add perturbatively an antiferromagnetic nearest-neighbor interaction $J_1$ between the two copies. In second order we arrive at an Ashkin-Teller model which is in agreement with a scenario of non-universal exponents.

The paper is structured as follows: we will present the model in detail in the following section before reviewing the MC simulations in Ref.~\onlinecite{P:kalz08} and new results that we will present in Sec.~\ref{s:MC}. We will also show that the length scales, which are needed to see the first-order features, are growing with $J_2$ and are not accessible any more with MC simulations for $J_2 \geq J_1$. The same scaling problems occur for the transfer matrix calculations in Sec.~\ref{s:tm}, which are used to determine the central charge of the underlying field theory in the limit of $J_2 \searrow J_1$. In Sec.~\ref{s:cf} we derive the conformal field theory for the case of two independent Ising models which interact via a perturbation caused by the nearest-neighbor coupling $J_1$.

\begin{figure}
\includegraphics[width=0.45\textwidth]{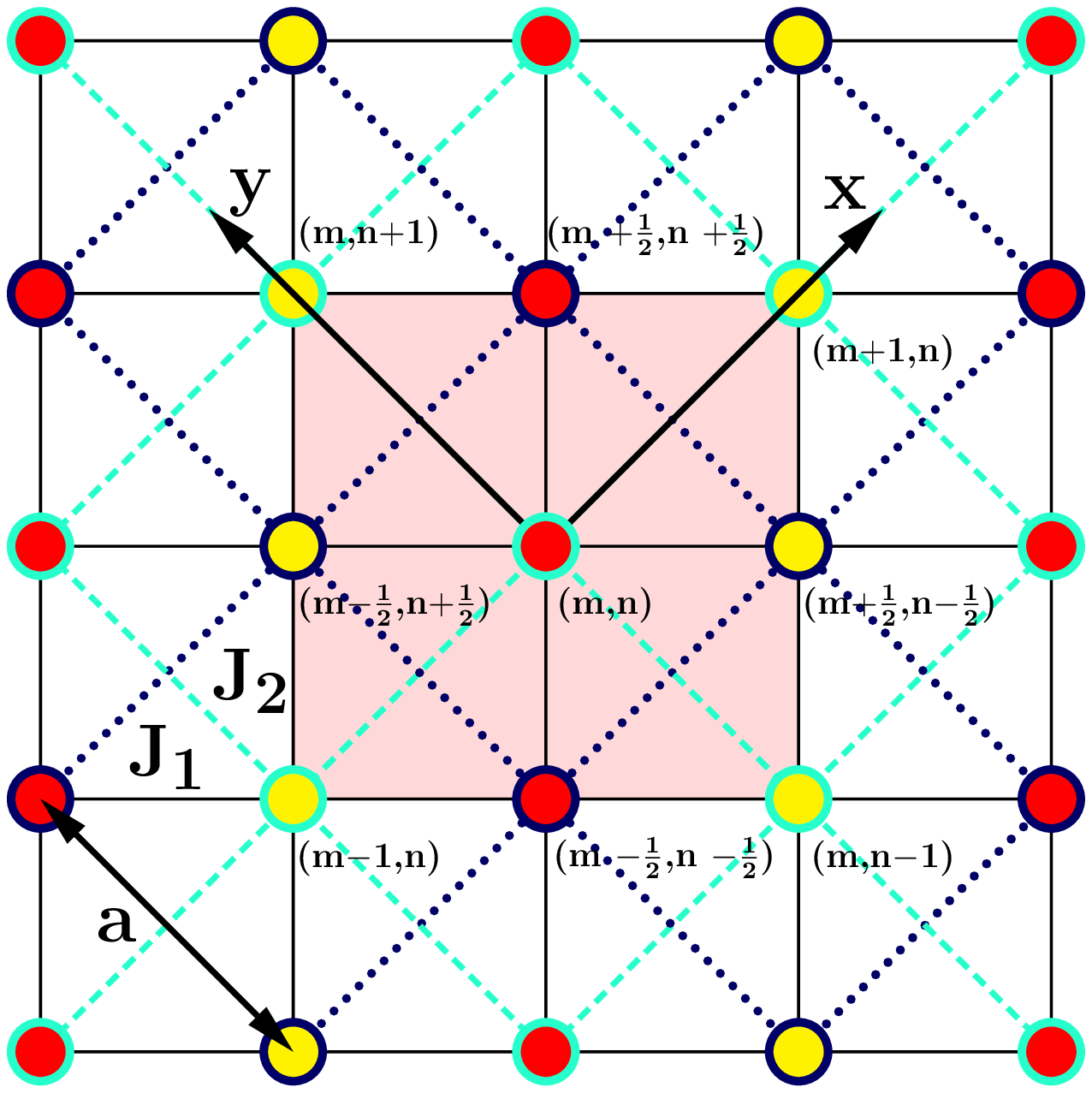}
\caption{\label{f:unit_a}(Color online) Collinear phase of the $J_1-J_2$ square lattice. Red dots stand for up spins, and yellow dots stand for down spins. The two copies $A$ and $B$ of the Ising model with magnetic couplings $J_2$ and lattice spacing $a$ are, respectively, represented with dashed clear and dotted dark blue lines, while the black thin lines correspond to the $J_1$ square lattice. The shaded area represents the unit cell used to derive the continuum limit (Sec.~\ref{s:continuum}). The coordinates are indicated with respect to the $x$- and $y$ axis of the $A$ sublattice.}
\end{figure}

\section{Model}\label{s:model}
\begin{figure}[h!]
\includegraphics[width=0.48\textwidth]{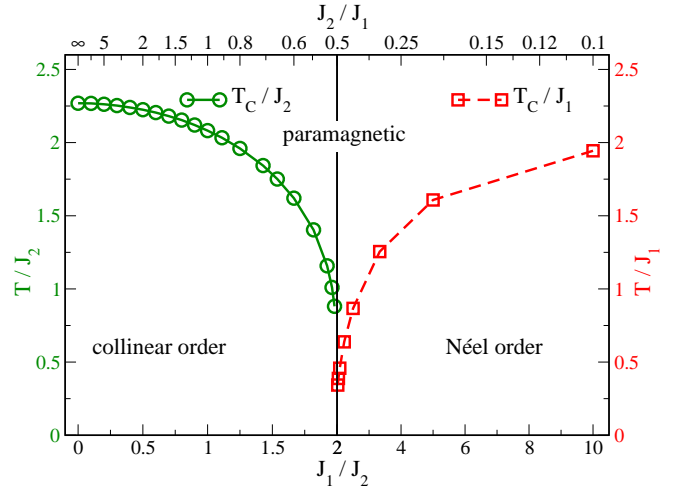}
\caption{\label{f:phase}(Color online) Critical temperatures for the phase transition from the paramagnetic into the magnetically ordered phase over the strength of frustration $J_1/J_2$. We adapted the energy scale of the temperature for $J_1/J_2 < 2$ to $J_2$ and for $J_1/J_2 > 2$ to $J_1$. Note that the frustration is also given in units of $J_2/J_1$ on the upper $x$-axis.}
\end{figure}

The lattice model is described by the Hamiltonian
\begin{align}
	\mathcal{H}_{\text{Ising}} = J_1 \sum_{\text{NN}} \si \cdot \sj + J_2 \sum_{\text{NNN}} \si \cdot \sj \label{e:hamil}\,,
\end{align}
where the sums run over all nearest-neighbor and next-nearest-neighbor interactions on a $N = L \times L$ square lattice (see Fig.~\ref{f:unit_a}). The energy on each bond is given by the product of the adjacent classical Ising spins $\si = \pm 1$ and the corresponding $J_i$ which are both chosen antiferromagnetic. Thus, a configuration which yields minimal energy for all bonds does not exist for finite $J_i$ and, hence, the model is frustrated. For small competing interactions $J_2 < J_1/2$ on the diagonals of the square lattice, the model undergoes a phase transition from the paramagnetic phase into a Néel ordered configuration at a critical temperature $T_C(J_2)$ which depends on the frustration (compare Fig.~\ref{f:phase}). This phase transition is continuous and the scaling exponents are the same as for the unfrustrated square lattice Ising model.\cite{B:lanbin00,P:kalz08}  If $J_1 = 2~J_2$ the critical temperature is suppressed to zero and the ground state is degenerate of order $2^{L}$.\cite{P:kalz08, P:kalz09} 
For smaller $J_1$ the ground state is a collinear antiferromagnet where lines of parallel spins are coupled anti-parallel.

\section{Monte Carlo simulation} \label{s:MC}
Based on the results we have presented in Refs.~\onlinecite{P:kalz08} and \onlinecite{P:kalz09} we performed further MC simulations on larger lattices and for new parameters. We used a single-spin Metropolis MC update\cite{P:metropolis53} with an additional line update. For an optimized thermalization process and high-data quality we also implemented an exchange Monte Carlo update\cite{P:hukushima96, P:hansmann97, P:katzgraber06} and ran the simulations on large-scale clusters using OpenMP and MPI. The statistical errors of the data were obtained by multiple independent MC runs.

\begin{figure}[h!]
\includegraphics[width=0.48\textwidth]{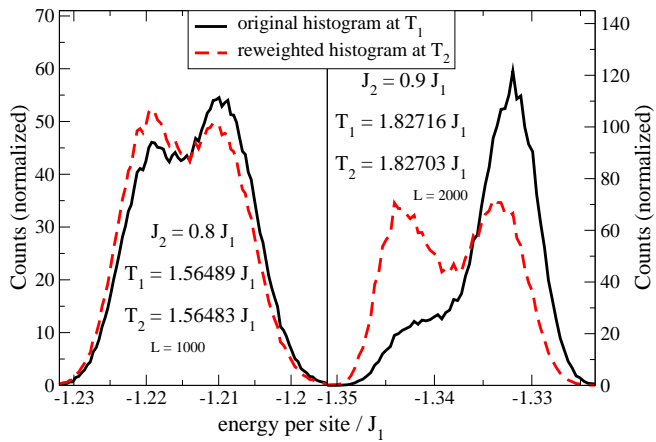}
\caption{\label{f:histo}(Color online) Energy histograms (solid black lines) for two values of $J_2 = 0.8~J_1$ (left) and $J_2=~0.9~J_1$ (right) for a $L = 1000$ (left) and $L=2000$ (right) lattice and the reweighted histograms for slightly lower temperatures (dashed red lines). The two-peaked structure emphasizes the first-order character of the phase transition.}
\end{figure}

The phase diagram ($T_C$ over $J_1/J_2$) showing both old data and new transition temperatures up to $J_2 = 10~J_1$ is given in Fig.~\ref{f:phase}. Additionally we looked at the character of the finite-temperature phase transitions and calculated critical exponents via finite-size scaling from our MC data. Comparing with old results from Landau and Binder \cite{P:lanbin80, P:landau80,P:lanbin85,B:lanbin00} we found a discrepancy between their values and ours for $J_2 > 0.5~J_1$. To have a closer look at the nature of the phase transition for this part of the phase diagram we recorded energy histograms at discrete temperature steps. These are plotted for $J_2 \leq 0.7~J_1$ in Fig.~4 of Ref.~\onlinecite{P:kalz09} and show a clear two-peaked structure and therefore prove the first-order character of the phase transition. Since López \et claim the first-order transition scenario to be valid up to $J_2 \simeq 1.1~J_1$,\cite{P:lopez93, P:lopez94} we recorded, for the present work, histograms also for larger values of $J_2$. In Fig.~\ref{f:histo} we show the recorded histograms of the MC simulations (as solid black lines) for a $2000 \times 2000$ lattice at $J_2 = 0.9~J_1$ and a $1000 \times 1000$ lattice at $J_2 = 0.8~J_1$. Note that for a small step in the parameter space ($0.1~J_1$) it is already necessary to double the linear size of the simulated lattice to achieve a similar resolution for the recorded histogram. The shapes of these histograms show a strong deviation from the almost Gaussian shape that is expected for a continuous phase transition, yet for the given temperatures the structure is not symmetric. For this reason, we also present reweighted histograms (as dashed red lines) for slightly lower temperatures, i.e., the size-dependent transition temperatures. The histograms are thus shifted to lower energies and exhibiting a more distinct and symmetric two-peak structure. The simulation and recording of a new histogram for this temperature would have been too time consuming and, hence, we applied the standard reweighting technique\cite{B:berg04, P:ghoufi08} to prove the first-order character of the phase transition.

\begin{figure}[h!]
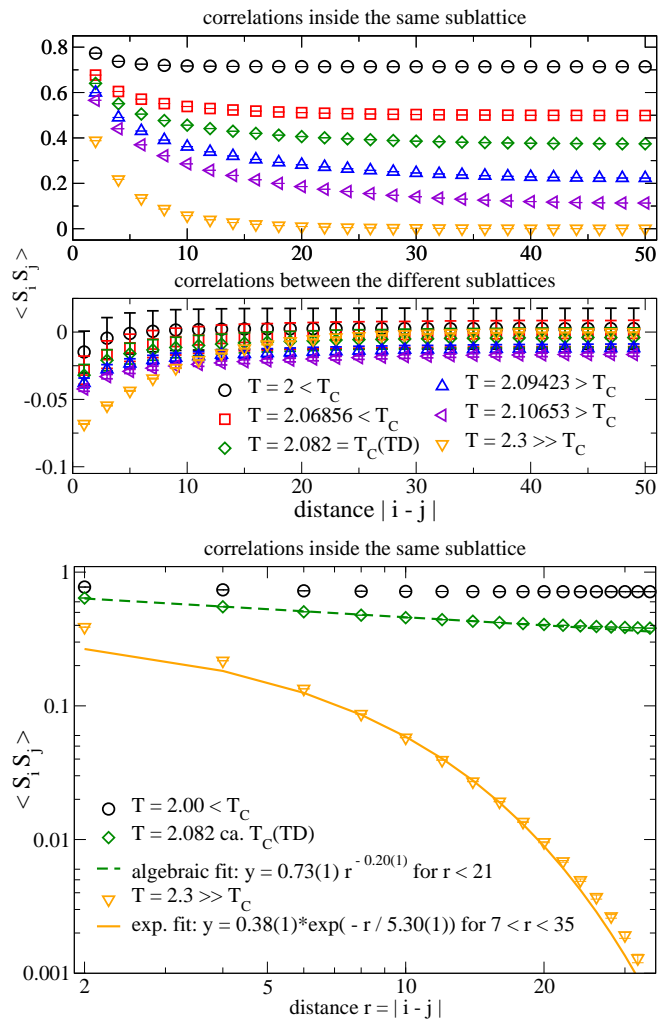

\subfigure{\includegraphics[width=0.48\textwidth]{corr_10_10.eps}}
\subfigure{\includegraphics[width=0.48\textwidth]{corr_10_10log.eps}}
\caption{\label{f:corr1}(Color online) Calculations for the spin-spin correlation functions $\langle S_i S_j \rangle$ at $J_1 = J_2$ for a $100 \times 100$ periodic lattice along one direction. Thus maximal distance is given by $i-j = 50$. \textbf{Top:} Correlations inside the same sublattice -- $i-j$ even -- for different temperatures around $T_C$. \textbf{Middle: } Correlations between spins on different sublattice sites -- $i-j$ odd --  in a larger scale. For all temperatures the correlations decay rapidly and go to zero. Hence, no long-range correlation is observable between the two sublattices. \textbf{Botttom: } Doubly logarithmic plot of the correlation functions inside the same sublattice for three exemplary temperatures and their related fits.}
\end{figure}

For larger $J_2$ the resolution of a double peaked profile in the histograms is not feasible, probably due to the growing crossover scales.

In addition we present measurements of the spin-spin correlation function $\langle S_i S_j \rangle$. In Fig.~\ref{f:corr1} we show separately the correlation functions $\langle S_i S_j \rangle$ for $i$ and $j$ being part of the same sublattice (top) and being in different sublattices (middle) for the value $J_1 = J_2$. For the correlations inside the same sublattice -- in addition shown in a doubly logarithmic scale (Fig.~\ref{f:corr1} bottom) -- we observe the behavior of a second-order phase transition, i.e., exponential decay for $T > T_C$, algebraic decay for $T \approx T_C$ and fast saturation toward a constant finite value for $T< T_C$. We also extracted the critical exponent $\eta$ which describes the scaling of the correlation in the vicinity of the critical temperature and obtained $\eta = 0.20(1)$. This is far away from the Ising value of $\eta_I = 0.25$. But for smaller values of $J_1/J_2$ the critical behavior becomes more Ising-like and we obtain values of $\eta = 0.25(1)$ for $J_1 / J_2 \leq 0.2$ (not shown). In the middle part of Fig.~\ref{f:corr1} we show the correlations between spins on different sublattices which decay for all temperatures quicker than correlations in the same sublattice -- note the different scales for the two upper panels of Fig.~\ref{f:corr1}. For $T < T_C$ the correlations drop to zero. We observed this behavior for all $J_1 / J_2 < 2$, i.e., in the region where the system undergoes a phase transition into the collinear phase. But the decay in $\langle S_i S_j\rangle$ becomes slower for increasing $J_1 /J_2$.

In conclusion the MC data yield a clear picture only for $0.5\,J_1 < J_2 \lesssim J_1$ where a first-order phase transition scenario is established by the doubly peaked structure of the energy histograms. For larger values of $J_2$ the analysis of the correlation functions indicates a decoupling of the two sublattices and a continuous phase transition. However, a detailed examination of the scaling behavior at the critical temperature and a reliable calculation of critical exponents are hampered by large crossover scales.

\section{Conformal field theory}\label{s:cf}
In the limit $J_1=0$, the system is exactly described by two decoupled two-dimensional Ising models on sublattices $A$ and $B$ (see Fig.~\ref{f:unit_a}),  and the critical behavior can therefore be tackled by conformal field theory.\cite{B:CFT_book, B:mussardo, Belavin_1984} A CFT is characterized by a constant $c$ called the central charge (or conformal anomaly). Various CFTs with $c<1$ were identified as statistical models at their critical point. In particular, the CFT with $c=1/2$, which was known to correspond to the massless free Majorana fermion, was identified as the critical two-dimensional Ising model.\cite{Belavin_1984, Belavin_1984_2}
Physically, the central charge characterizes the short-distance behavior of the theory and can be seen as a measure of the number of degrees of freedom of the system. Central charges of decoupled systems add up and, therefore, in the limit $J_1=0$ where the two Ising models $A$ and $B$ are independent, it is $c=1$. At this point we want to emphasize that every unitary theory with $c<1$ yields a universality class with constant universal exponents. Hence, to gain a phase transition with varying critical exponents is only possible for a CFT with $c\geq 1$.\cite{Ginsparg_1988}

\subsection{Transfer Matrix}\label{s:tm}
In the 1980s it was shown that, for a lattice model, the free energy $f$ per site of a cylinder of infinite length and finite circumference $L$ shows a finite-size scaling in $1/L$ with a proportionality factor depending on the central charge of the corresponding field theory.\cite{P:cardy86,P:affleck86}
\begin{align}
f = a - c \frac{\pi\,T_C}{6\,L^2} + \mathcal{O}(L^{-4})\,,\quad a =\text{const.}
\end{align}
To calculate the free energy of our model, MC simulations are not suitable since the entropy is not available for them. Therefore, we chose to implement a transfer-matrix algorithm.\cite{P:nijs82, B:baxter82} We were able to calculate the free energy for systems of size $L \times B$, where the circumference of the cylinder is limited to $L \leq 22$ because of exponential growth of computational effort and the length can be chosen easily up to $B = 10000$ (linear scale). A finite-size scaling of the free energy yields an estimate of the central charge. We present our results for different $J_2/J_1$ in Table~\ref{t:cc}. For large values of $J_2$ the central charge seems to converge to the value of two independent Ising models, $c=1$. However, for decreasing $J_2$ they do not converge and we find $c > 1$, which is not allowed by the $c$ theorem which states that the central charge can not increase under the influence of a renormalization-group transformation if the corresponding field theory is critical.\cite{Zamalodchikov_1986} Indeed the MC data indicates a weak first-order transition for $J_2 = 0.6\,J_1$\cite{P:kalz08} and $J_2 = 0.8\,J_1$ (Fig.~\ref{f:histo} left-hand side) accompanied by large crossover scales. Thus, it is not surprising that this weak first-order transition is not detected by the transfer matrix computations for cylinders with circumferences $L \leq 22$ such that the corresponding results for $c$ are not meaningful.

\begin{table}
\begin{tabular}{cccc}
\hline \hline
$~J_2 / J_1~$	&	$c$ & $~J_2 / J_1~$	&	$c$ \\
\hline 
$~0.0~$	& $0.4999(1)$ & $~1.0~$ & $1.0613(6)$ \\
$~0.2~$	& $0.4994(3)$ &  $~1.5~$ & $1.0206(2)$ \\
$~0.6~$	& $(1.5811(18))$ &  $~2.5~$ & $1.0062(4)$ \\
$~0.8~$ & $(1.1273(10))$ &  $~10~$ & $1.0000(4)$ \\
\hline \hline
\end{tabular}
\caption{\label{t:cc}Central charge of the underlying field theory for different $J_2 / J_1$ calculated with a transfer matrix computation of the free energy. The width of the computed systems satisfies $L \leq 22$. We also include in parentheses $c$-values for $J_1/2 < J_2 < J_1$ where the phase transition is of weak first order according to the MC analysis.}
\end{table}

\subsection{Continuum limit}\label{s:continuum}
In order to get more information on the nature of the phase transition for $J_1 \ne 0$ we now derive our model in the continuum limit. We start from the conformally invariant fixed point $J_1=0$ (see Fig.~\ref{f:unit_a}) and then add perturbatively a coupling $\propto J_1$ between the two decoupled Ising models $A$ and $B$. 

In a first step, the spin variables on discrete lattice sites are replaced by continuous fields which resemble the underlying Néel order of the two decoupled antiferromagnetic Ising models. Thus, the staggered spin variables $\mathbf{S}_I$ ($I= A,B$ sublattice) have to be transformed into smooth variables defined by
\be
 \sigma_I(m,n) \propto (-1)^{m+n}\mathbf{S}_I(m,n)\,,
 \label{e:gauge}
\ee
where $(m,n)$ are the lattice coordinates. Note that the transformation Eq.~\eqref{e:gauge} is based on a specific choice of gauge but this does not affect the macroscopic properties of the system. The choice of the unit cell is shown in Fig.~\ref{f:unit_a} and the coordinate system is rotated by an angle of $\pi/4$; i.e., the axes point along the next-nearest neighbor bonds. In a next step the sum of the Hamiltonian equation~\eqref{e:hamil} is converted into a two-dimensional integral where the values of $\sigma_I (x,y)$ at the limits $(-\infty, \infty)$ are equal due to the periodic boundary conditions imposed on our model:
\be
\sum_{\text{i,j}} \rightarrow \frac{1}{a^2}\int_{-\infty}^{\infty}dx dy\,,\quad a \text{: lattice spacing}
\ee
A Taylor expansion up to second order on the $\sigma_I = \sigma_I(x,y)$ fields is given by:
\bea
&&\sigma_I(x+ m\,a,y+ n\,a) = \sigma_I + a\left[m\partial_x+ n\partial_y\right]\sigma_I \nonumber \\
&&+ \frac{a^2}{2}\left[m^2\partial^2_{xx}+n^2\partial^2_{yy}+2mn\partial^2_{xy}\right]\sigma_I +\mathcal{O}(\partial^3 \sigma_I)\,.
\eea
By summing up all nearest-neighbor interactions ($J_1$) appearing in the chosen unit cell, the products $\sigma_A \sigma_B$ and $\sigma_I \partial_{x,y} \sigma_J$ ($I\neq J$) are canceled out due to frustration.\footnote{This was overlooked in an earlier discussion by two of the authors in Ref.~\onlinecite{P:kalz08}.} This feature distinguishes the present model from the two-layer Ising model, where the highly relevant $\sigma_A\sigma_B$ coupling survives.\cite{B:delfino98, P:simon00} The $J_1$ interaction is then given by:
\bea
	\mathcal{H} _{\text{int}} &= &-\mu^2 J_1 \iint dxdy \Big( \p_x\sigma_A\p_y\sigma_B + \p_y\sigma_A\p_x\sigma_B \Big)\,. \label{eq:Interaction_continuum}
\eea
Note that partial derivatives are understood to act only on the subsequent operator and $\mu$ is a constant factor.

For further calculations, it is convenient to rewrite the interaction [Eq.~\eqref{eq:Interaction_continuum}] in complex coordinates $z=x + i y$ and $\bz=x - iy$:
\bea
 \mathcal{H}_{\text{int}} &=& - i \mu^2 J_1 \miintz \Big(\mathcal{O}_1(z,\bz) - \mathcal{O}_2(z,\bz) \Big) 
 \label{eq:Interaction_complex}\\
 \mathcal{O}_1(z,\bz)&=&  \p_z\sigma_A\p_z\sigma_B \,\, , \,\, \mathcal{O}_2(z,\bz)=  \pbz\sigma_A\pbz\sigma_B\,.
 \label{eq:Interaction_operators} 
 \eea
The $\sigma_I$ fields of the theory have conformal dimensions $(h_{\sigma},\bar{h}_{\sigma})=(1/16,1/16)$.\cite{Belavin_1984, Ginsparg_1988, B:CFT_book, B:mussardo} One defines the scaling dimension $\Delta=h+\bar{h}$ and the conformal spin $s=h-\bar{h}$. 
The full model is then described by the action
\begin{align}
\mathcal{A} &= \mathcal{A}_{A}^{0}+\mathcal{A}_{B}^{0}+\tau \miintz \Big(\varepsilon_A(z,\bz) +\varepsilon_B(z,\bz)\Big)\label{e:pure_action} \\
 +& ~g \miintz \Big(\mathcal{O}_1(z,\bz) - \mathcal{O}_2(z,\bz) \Big)\,,\quad g \propto J_1\,, \label{e:full_action}
 \end{align}
 where $\mathcal{A}_I^0$ are the fixed-point actions of the Ising models $A$ and $B$. Furthermore $\varepsilon_I$ are the usual thermal operators of the conformal field theory on the two-dimensional Ising model with $(h_{\varepsilon}, \bar{h}_{\varepsilon})=(1/2,1/2)$ and the corresponding mass $\tau \propto (T-T_C^{J_1=0})$. 
 
The operators $\mathcal{O}_1$ and $\mathcal{O}_2$ are kept, despite being highly irrelevant with a scaling dimension $\Delta = 9/4$, due to their non-zero conformal spins $s = \pm 2$. The presence of such chiral terms was previously reported in other frustrated systems such as the anisotropic square lattice,\cite{Starykh_2004} the checkerboard \cite{Starykh_2005} and the Kagom\'e \cite{Schnyder_2008} lattices. These \emph{twist terms}\cite{Nersesyan_1998} are known to be likely to generate relevant or marginally relevant terms at higher order.\cite{Tsvelik_2001, Allen_2001}

Before calculating higher orders of the perturbative interaction, we want to briefly discuss the underlying symmetries of the model and the consequences for the continuous field theory.
The Hamiltonian Eq.~\eqref{e:hamil} is invariant under translations by multiples of the lattice spacing $a$, which is preserved for the integral form of the field theory because of the same periodic boundary conditions imposed on the integrals. Furthermore, the model is symmetric under rotations by an angle of $\pi/2$ and inversions along the axes $(x,y=0)$, $(x=0,y)$, $(x,y=x)$ and $(x,y=-x)$ (corresponding to the diagonals and the vertical and horizontal line through the origin in Fig.~\ref{f:unit_a} residing on an $A$ site of the lattice). Note that the rotations and inversions are not independent. In addition the model is symmetric under the exchange of the two sublattices, which is equivalent to the translation by one lattice spacing of the original lattice. The inversion, e.g., of the axis $(x,y=0) \rightarrow (-x,y=0)$ is given in the complex coordinates by setting:
\begin{align}
&z \rightarrow -\bar z,\, \bar z \rightarrow - z, \, \partial_z \rightarrow -\partial_{\bar z},\, \partial_{\bar z} \rightarrow -\partial_{z},\nonumber\\ 
& \sigma_A (z,\bar z) \rightarrow \sigma_A (-\bar z, -z) \text{ and }\nonumber\\
&\sigma_B (z,\bar z) \rightarrow -\sigma_B (-\bar z, -z) \label{e:inver_x}
\end{align}
Since the $\sigma_I$ fields are completely symmetric in $z$ and $\bar z$ and we integrate over the whole complex plane for both variables, the crucial point of this transformation is the change of the partial derivatives and the additional sign that occurs for the $B$ sublattice. Thus, operators that contribute to the continuous field theory have to be either quadratic in the sublattice fields or contain an asymmetric contribution of derivatives and fields residing on different sublattices, as those in Eq.~\eqref{eq:Interaction_operators}.
Furthermore, the inversion $(x,y=-x)\rightarrow (-x,y=x)$ rotates the partial derivatives onto the imaginary axis, which ensures that by symmetry only operators containing an even number of partial derivatives are allowed. Following these symmetry arguments, we can discuss the appearance of certain operators in higher orders. A highly relevant ($\Delta = 1/4$) spin-spin coupling $\sigma_A(z,\bar z)\sigma_B(z,\bar z)$ such as the one appearing in the two-layer Ising model\cite{B:delfino98, P:simon00} is not allowed, whereas combinations of energy operators $\varepsilon_I$ are allowed since they transform like products of spin fields on the same sublattice. 

\subsection{Operator product expansion}\label{s:OPE}
Higher-order perturbations are calculated via the standard operator product expansion (OPE).\cite{B:CFT_book, B:mussardo, Belavin_1984} This operation allows us to replace, inside a correlation function, a product of two operators by a combination of scaling operators allowed by the theory. This is meant to close the renormalization-group equations in the operator algebra of the model before discarding irrelevant perturbations. For the two-dimensional Ising model, the field content of the product of two fields is encoded in the fusion rules:
\bea
\left[\sigma_I\right] \left[\sigma_J\right] &=& \delta_{I,J} \big( [\identity] + [\varepsilon_I] \big) \nn
 \left[\varepsilon_I\right] \left[\varepsilon_J \right] &=& \delta_{I,J} [\identity]  \nn
 \left[\sigma_I\right] \left[\varepsilon_J\right] &=& \delta_{I,J} [\sigma_I]
 \label{eq:Fusion_rules} \,,
\eea
where  $\identity$ is the identity operator present in all CFTs and $I,J$ represent the two sublattices $A$ and $B$.
Using the general normalized form of an OPE and the fusion rules, one obtains:\cite{B:CFT_book}
\bea
&\sigma_I(z,\bar z)\sigma_J(w,\bar w) = \frac{\delta_{I,J}}{|z-w|^{1/4}} + \frac{\delta_{I,J}}{2}|z-w|^{3/4} \varepsilon_I(w,\bar w) \nonumber\\
&+ \frac{\delta_{I,J}}{4}(z-w)^{11/8}(\bz-\bw)^{3/8} \p_w \epsilon_I(w,\bw)+\text{H.c.} \,.
\label{e:OPE_sigma_sigma}
\eea
Note that we kept marginal terms with non-zero conformal spin to be consistent with previous comments about the relation of the eventual importance of chiral terms in frustrated systems.

By calculating the OPE between the $\mathcal{O}_{1,2}$ operators of Eq.~\eqref{eq:Interaction_operators} one generates higher-order terms. At second order this generates terms [$\varepsilon = \varepsilon (w,\bar w)$]:
\begin{align}
\propto~&~ \varepsilon_I \,,\quad \propto~ \partial_{\{w,\bar{w}\}} \varepsilon_I\,,\quad \propto~\varepsilon_I \varepsilon_J \label{e:rel}\\
\propto~& \varepsilon_I \partial_{\{w,\bar{w}\}} \varepsilon_J \, \quad \text{and} \quad \propto~\partial_{\{w,\bar{w}\}} \varepsilon_I\partial_{\{w,\bar{w}\}} \varepsilon_J \quad \scriptstyle(I\neq J)\label{e:irrel}\,.
\end{align}
 Following the calculations presented in Appendix we get: \footnote{Note that for symmetry reasons we calculate the OPE of the full interaction with itself.}
\begin{widetext}
\begin{align}
&\left[\mathcal{O}_{1}(z,\bar z)-\mathcal{O}_{2}(z,\bar z)\right]\left[\mathcal{O}_{1}(w,\bar w)-\mathcal{O}_{2}(w,\bar w)\right] =
\alpha_0\left[(z-w)^{-\frac{17}{4}}(\bar z - \bar w)^{-\frac{1}{4}}+\text{H.c.}\right]-\frac{2}{4096}|z-w|^{-\frac{9}{2}} \nonumber \\
&- \left\{ \alpha_1\left[(z-w)^{-\frac{15}{4}}(\bar z - \bar w)^{\frac{1}{4}}+\text{H.c.}\right]
-\frac{9}{4096}|z-w|^{-\frac{7}{2}}\right\}(\varepsilon_A + \varepsilon_B)\label{e:alpha} \\
&+ \left\{\alpha_2\left[(z-w)^{-\frac{13}{4}}(\bar z -\bar w)^{\frac{3}{4}}+\text{H.c.}\right] 
- \frac{81}{8192}|z-w|^{-\frac{5}{2}}\right\}(\varepsilon_A\varepsilon_B)\label{e:gamma} \\
&- \left(\alpha_3(z-w)^{-\frac{3}{4}}(\bar z - \bar w)^{-\frac{7}{4}}+\alpha_4(z-w)^{-\frac{11}{4}}(\bar z - \bar w)^{\frac{1}{4}} +\alpha_3(z-w)^{\frac{5}{4}}(\bar z - \bar w)^{-\frac{15}{4}} \right) (\partial_w \varepsilon_A+\partial_w \varepsilon_B) + \text{H.c.} \label{e:beta1}\\
&+ \left(\alpha_5(z-w)^{-\frac{9}{4}}(\bar z -\bar w)^{\frac{3}{4}}-\alpha_6(z-w)^{-\frac{1}{4}}(\bar z -\bar w)^{-\frac{5}{4}}+\alpha_5(z-w)^{\frac{7}{4}}(\bar z -\bar w)^{-\frac{13}{4}}\right)(\partial_w \varepsilon_A\varepsilon_B+\varepsilon_A\partial_w\varepsilon_B) + \text{H.c.}\label{e:irrel1} \\
&+\sum_{k} \beta_k (z-w)^{v_k} (\bar z -\bar w)^{t_k}\big(\mathcal{O}(\partial_{\{w,\bar{w}\}}^2\varepsilon_A\varepsilon_B)\big)\,,\quad v_k \neq t_k \label{e:irrel2}
\end{align}
\end{widetext}
The coefficients $\alpha_i$ and $\beta_k$ are some rational constants, and the terms in Eq.~\eqref{e:irrel1} are irrelevant but could -- as shown before -- produce again relevant terms in higher order. All terms in Eq.~\eqref{e:irrel2} are also irrelevant and contain second derivatives that will produce only highly irrelevant terms ($\Delta=4$) in higher orders.

The third-order $\propto J_1^3$ terms are obtained by multiplying the above operators with the original perturbation Eq.~\eqref{eq:Interaction_operators}. One needs the following OPE:
\bea
&\sigma_I(z,\bz) \varepsilon_J(w,\bw) = \frac{\delta_{I,J}}{2}|z-w|^{-1}\sigma_I(w,\bw) \label{eq:sig_eps_OPE}\nonumber \\
+& \frac{\delta_{I,J}}{4}(z-w)^{\frac{1}{2}}(\bar z -\bar w)^{-\frac{1}{2}}\partial_{w}\sigma_I(w,\bw) + \text{H.c.}
\eea
It yields only three different types of operators that are primary and secondary operators from the spin family:
\bea
\propto& ~\sigma_I\sigma_J \,, \quad \propto~\sigma_I \partial_{\{w,\bar{w}\}}\sigma_J \label{e:thirdrel}\\
&\text{and} \quad \propto~\partial_{\{w,\bar{w}\}}\sigma_I \partial_{\{w,\bar{w}\}}\sigma_J \,. \label{e:thirdirr}
\eea
Calculating the third-order perturbation by multiplying every operator from the second order with $\mathcal{O}_{1}-\mathcal{O}_{2}$ yields the prefactors for these spin operators. However, for each of the relevant operators in Eq.~\eqref{e:thirdrel} the sum of all prefactors gives exactly zero, which is in agreement with the symmetry considerations presented in Sec.~\ref{s:continuum}. Thus, the third order does not give any new operators since the irrelevant terms in Eq.~\eqref{e:thirdirr} are the same as in the first order of our perturbation.  Thus, we have closed the operator algebra and are left only with the terms of Eq.~\eqref{e:rel}, namely, the thermal operators $\varepsilon_A + \varepsilon_B$ and two marginal operators, $\varepsilon_A\varepsilon_B$ and $\p_{\{w,\bar{w}\}}\varepsilon_A + \p_{\{w,\bar{w}\}}\varepsilon_B$. 

The next step is the integration of the prefactors given in Eqs.~\eqref{e:alpha}, \eqref{e:gamma} and \eqref{e:beta1} which depend on the product of $(z-w)^{v}$ and $(\bar z - \bar w)^{t}$. Since for the second-order terms in Eq.~\eqref{e:rel} we have to integrate over all four variables, $z,~\bar z,~w$, and $\bar w$, with the only constraint being $|z-w| > a$, all prefactors with exponents $v \neq t$ and $|v-t| =n$ (with $n \in \mathbb N$) will be zero due to the phase integration from $0$ to $2\pi$. 
In detail one can carry out first the integral over $z$ and $\bar z$, which leaves the energy fields untouched; remembering that $z$ and $\bar z$ are complex conjugates, one can rewrite $(z-w)^{v}(\bar z - \bar w)^{t} = |z-w|^{v+t}e^{i\phi (v-t)}$, whereby $\phi = \arg (z-w)$.
 
This integration cancels the marginal chiral terms $\partial_w \varepsilon_I$ and $\partial_{\bar w} \varepsilon_I$, again in agreement with the symmetries of the model. Thus, the perturbation is given by the pure energy terms and the energy-energy coupling
\begin{align}
-\frac{9\mu^4 J_1^2}{4096}&\int_{|z-w|>a}dzd\bar z dw d\bw~|z-w|^{-\frac{7}{2}}(\varepsilon_A+\varepsilon_B)\label{e:AT_sum} \\
+\frac{81\mu^4 J_1^2}{8192}&\int_{|z-w|>a}dzd\bar z dw d\bw~|z-w|^{-\frac{5}{2}}\varepsilon_A \varepsilon_B\label{e:AT_product}\,.
\end{align}
The signs of these operators stem from the squares of the coupling given in Eq.~\eqref{eq:Interaction_complex} and the derived signs of the perturbative calculations given in Eqs.~\eqref{e:alpha} and \eqref{e:gamma}, respectively. The pure energy terms in Eq.~\eqref{e:AT_sum} and the mass term in Eq.~\eqref{e:pure_action} are necessary to tune the model onto criticality. We find the new critical temperature up to second order in perturbation theory
\bea
T_C(J_1) = T_C(0) - \gamma ~J_1^2\,,
\eea
which appears to be in agreement with the behavior of the critical line on the left-hand side of the phase diagram presented in Fig.~\ref{f:phase}. The rescaled mass term is equal to zero on the critical line, thus the most relevant perturbation is the marginal energy-energy coupling [Eq.~\eqref{e:AT_product}]. Together with the action of the unperturbed Ising models from Eq.~\eqref{e:pure_action} the \emph{Ashkin-Teller} field theory is readily identified:\cite{B:Ashkin_1943,B:delfino98}
\begin{align}
\mathcal{A}_{AT} = \mathcal{A}_{A}^{0}+\mathcal{A}_{B}^{0} + k \int dw d\bw ~\varepsilon_A \varepsilon_B \label{e:AT_action}
\end{align}
where $k \propto J_1^2$ is the renormalized coupling constant. The sign of the marginal perturbation can in principle matter, but for the present case it does not affect the conclusion that we stay on the critical line. Since the free fermionic theory is located in the middle of a line of $c=1$ conformal field theories (see, for example, Ref.~\onlinecite{Ginsparg_1988}), the
theory is critical on either side of the fixed point of two decoupled Ising models.

\subsection{Ashkin-Teller model}
The Ashkin–Teller lattice model was introduced as a generalization of the Ising model to a four-component system.\cite{B:Ashkin_1943} However, in the 1970s it was shown that the model can be mapped onto a system of two Ising models ($A$ and $B$) residing on the same lattice and interacting via an additional four-spin interaction:\cite{P:fan72}
\begin{align}
 \mathcal{H}_{AT} &= J \!\!\!\!\!\!\! \sum_{\text{NN} \in \mu={A,B}} \!\!\!\!\!\!\! \si_{\mu} \cdot \sj_{\mu} 
		   + J_4 \sum_{\text{NN}} \si_A \cdot \sj_A \cdot \si_B \cdot \sj_B
 \label{e:Ashkin-Teller} \,.
\end{align}
The field-theoretic action of this model is the same as in Eq.~\eqref{e:AT_action} with a coupling constant $k=f(J, J_4)$.
Thus, the four-spin interaction is mimicked by the perturbative onset of the energy-energy coupling, which is, in our case, proportional to the square of the original nearest-neighbor coupling $J_1$. The rich phase diagram of the Ashkin-Teller model given in coupling constants $J$ and $J_4$ in Refs.~\onlinecite{P:ditzian80, B:baxter82, P:Delfino04} includes a critical line which represents the one-dimensional flow diagram of the corresponding field theory [Eq.~\eqref{e:AT_action}] at criticality with a single parameter given by the renormalized coupling $k$. It starts for $k=0$ at the conformally invariant fixed point of two decoupled Ising models and ends at the Potts-critical end point. In our calculations we arrived at a CFT with $c=1$ in the presence of a marginal operator. Both are necessary conditions to fulfill so that the theory can exhibit varying critical exponents.\cite{B:Kadanoff71, B:Friedan1984} This scenario would be consistent with our numerical results for large $J_2$ and earlier descriptions of the frustrated Ising model.\cite{B:lanbin00} The Potts-critical end point would also allow for an onset of a non-critical line of first-order phase transitions as we see it in the MC simulations. However, since our effective field theory is only derived perturbatively, the critical behavior does not necessarily need to be described by the Ashkin-Teller model up to $ J_1 \approx J_2$. Thus, the explanation of the observed first-order transition ($1 \lesssim J_1/J_2 < 2$) might be beyond the scope of the Ashkin-Teller model.

\section{Discussion}
Motivated by Refs.~\onlinecite{P:lopez93, P:malakis06, P:kalz08, P:kalz09} we had a closer look at the phase transition from the paramagnetic phase into the antiferromagnetic collinear ordered phase of the two-dimensional frustrated $J_1$-$J_2$ Ising model. We performed extensive additional MC simulations to verify the weak first-order character of the transition for the particular values of $J_2/J_1 = 0.8$ and $0.9$ and recorded spin-spin correlation functions for larger values of $J_2/J_1$ to gain an insight into the phase transition. However, the increasing length scales in the system do not allow for a reliable interpretation of the phase transition for $J_1 \lesssim J_2$. Since for $J_1 = 0$ the model can be described by two decoupled Ising models and, hence, is treatable by means of CFT, we tried in a first attempt to calculate the corresponding central charge $c$ with respect to the degree of frustration $J_1/J_2$ using transfer-matrix techniques. This computation yielded the reasonable result $c=1$ for small intercoupling values $J_1$ but suffered the same scaling problems as the MC simulations for larger values of $J_1/J_2$. Thus, we derived the continuous field theory for the discrete lattice model starting at the point $J_1=0$ and computed the perturbation induced by the nearest-neighbor coupling. Using OPE we closed the renormalization-group equations and arrived at an effective action which resembles the Ashkin-Teller model with $c=1$. We have given symmetry arguments that highly relevant terms such as the spin-spin coupling operator can not be generated at any order. Thus, the critical behavior differs drastically from that of a two-layer Ising model.

The derived effective field theory has a marginal energy-energy coupling as the most relevant perturbation to the decoupled Ising fixed-point action. Since this action equals the one of the Ashkin-Teller model we arrived at a $c=1$ CFT which together with the presence of a marginal operator allows for continuous phase transitions with varying critical exponents.\cite{B:Kadanoff71, B:Friedan1984} Thus, the weak-universality scenario of Landau and Binder\cite{B:lanbin00} is recovered for $J_1 \ll J_2$ but rather explained by the Ashkin-Teller model which opens the possibility for the occurrence of the first-order phase transition we observed in the MC simulations. The fact that the Ashkin-Teller model is only reached in the second order of the perturbation explains the large length scales that we find in the numerical analysis of the problem: the direct continuum limit yields only the irrelevant operators [Eq.~\eqref{eq:Interaction_operators}]. However, while these operators flow to zero under the renormalization group, they drive the marginal Ashkin-Teller coupling [Eq.~\eqref{e:gamma}] to a finite value, thus generating a crossover behavior as a function of the length scale. Moreover, it suggests that a further numerical analysis of the critical behavior would need very large systems sizes.

It is also possible that the onset of a first-order transition is caused by a level crossing of higher-energy states rather than by the renormalization flow of the Ashkin-Teller model itself.

Our results for the nature of the finite-temperature phase transition should also be relevant if small quantum fluctuations are included.\cite{P:roscilde04, P:kalz11} Indeed Ref.~\onlinecite{P:roscilde04} also finds a first-order transition close to the highly frustrated point which turns into a second-order transition for large $J_2$ in a certain quantum generalization of the present model.

\begin{acknowledgments}
We would like to give special thanks to Philippe Lecheminant for fruitful discussions and careful analysis of our field theoretical arguments. We also thank Alexei Tsvelik for his time and helpful insights into conformal field theory.  Most of the simulations in this paper were performed on the clusters of the Gesellschaft für wissenschaftliche Datenverarbeitung Göttingen and we want to thank them for technical support. Furthermore we would like to thank the Deutsche Forschungsgemeinschaft for financial support via the collaborative research center SFB 602 (TP A18) and a Heisenberg fellowship (Grant No.~2325/4-2, A. Honecker).
\end{acknowledgments}

\begin{widetext}
\begin{appendix}
\section{Second order perturbation using OPE\label{app:ope}}
The calculation of the second-order terms given in Eqn.~\eqref{e:alpha}-\eqref{e:irrel2} is shown exemplarily for the terms in Eq.~\eqref{e:alpha} and \eqref{e:gamma}:
\begin{align}
&\left[\mathcal{O}_{1}(z,\bar z)-\mathcal{O}_{2}(z,\bar z)\right]\left[\mathcal{O}_{1}(w,\bar w)-\mathcal{O}_{2}(w,\bar w)\right] \nn
=& \mathcal{O}_{1}(z,\bar z)\mathcal{O}_{1}(w,\bar w)-\mathcal{O}_{1}(z,\bar z)\mathcal{O}_{2}(w,\bar w)-\mathcal{O}_{2}(z,\bar z)\mathcal{O}_{1}(w,\bar w)+\mathcal{O}_{2}(z,\bar z)\mathcal{O}_{2}(w,\bar w) \\
=& \p_z\sigma_A(z,\bz)\p_z\sigma_B(z,\bz)~\p_w\sigma_A(w,\bw)\p_w\sigma_B(w,\bw) - \p_z\sigma_A(z,\bz)\p_z\sigma_B(z,\bz)~\pbw\sigma_A(w,\bw)\pbw\sigma_B(w,\bw) \nn
&- \pbz\sigma_A(z,\bz)\pbz\sigma_B(z,\bz)~\p_w\sigma_A(w,\bw)\p_w\sigma_B(w,\bw) + \pbz\sigma_A(z,\bz)\pbz\sigma_B(z,\bz)~\pbw\sigma_A(w,\bw)\pbw\sigma_B(w,\bw) \,.
\end{align}
Using the fact that only products of operators on the same sublattice are non-zero and extracting the partial derivatives, we rewrite the last line before applying the general normalized OPE [given in Eq.~\eqref{e:OPE_sigma_sigma}] for the spin-operator products:
\begin{align}
=& \p_z\p_w\sigma_A(z,\bz)\sigma_A(w,\bw)~\p_z\p_w\sigma_B(z,\bz)\sigma_B(w,\bw) - \p_z\pbw\sigma_A(z,\bz)\sigma_A(w,\bw)~\p_z\pbw\sigma_B(z,\bz)\sigma_B(w,\bw) \nn
&- \pbz\p_w\sigma_A(z,\bz)\sigma_A(w,\bw)~\pbz\p_w\sigma_B(z,\bz)\sigma_B(w,\bw) + \pbz\pbw\sigma_A(z,\bz)\sigma_A(w,\bw)~\pbz\pbw\sigma_B(z,\bz)\sigma_B(w,\bw) \\
=& \p_z\p_w\left( |z-w|^{-\frac{1}{4}}+\tfrac{1}{2}|z-w|^{\frac{3}{4}}\varepsilon_A(w,\bw)
+\tfrac{1}{4}(z-w)^{\frac{11}{8}}(\bz - \bw)^{\frac{3}{8}}\p_w\varepsilon_A(w,\bw)+\text{H.c.}\right)\nn
&\times \p_z\p_w\left( |z-w|^{-\frac{1}{4}}+\tfrac{1}{2}|z-w|^{\frac{3}{4}}\varepsilon_B(w,\bw)
+\tfrac{1}{4}(z-w)^{\frac{11}{8}}(\bz - \bw)^{\frac{3}{8}}\p_w\varepsilon_B(w,\bw)+\text{H.c.}\right)\nn
&-\p_z\pbw(\dots_A)\p_z\pbw(\dots_B)-\pbz\p_w(\dots_A)\pbz\p_w(\dots_B)+\pbz\pbw(\dots_A)\pbz\pbw(\dots_B) \,.
\end{align}
In the following we only regard the regular terms and the pure energy terms for the sake of clarity. Note that $|z-w|^x = (z-w)^{\frac{x}{2}}(\bz - \bw)^{\frac{x}{2}}$, which enters into the calculation of the partial derivatives.
\begin{align}
=& \left(-\tfrac{9}{64}(z-w)^{-\frac{17}{8}}(\bz-\bw)^{-\frac{1}{8}}
+\tfrac{15}{128}(z-w)^{-\frac{13}{8}}(\bz-\bw)^{\frac{3}{8}}\varepsilon_A(w,\bw)+\mathcal{O}(\p_{\{w,\bar{w}\}} \varepsilon_A)\right) \nn
&\times \left(-\tfrac{9}{64}(z-w)^{-\frac{17}{8}}(\bz-\bw)^{-\frac{1}{8}} \label{ae:o1o1}
+\tfrac{15}{128}(z-w)^{-\frac{13}{8}}(\bz-\bw)^{\frac{3}{8}}\varepsilon_B(w,\bw)+\mathcal{O}(\p_{\{w,\bar{w}\}} \varepsilon_B)\right) \\
&- 2\left(-\tfrac{1}{64}(z-w)^{-\frac{9}{8}}(\bz-\bw)^{-\frac{9}{8}}
+\tfrac{9}{128}(z-w)^{-\frac{5}{8}}(\bz-\bw)^{-\frac{5}{8}}\varepsilon_A(w,\bw)+\mathcal{O}(\p_{\{w,\bar{w}\}} \varepsilon_A)\right)\nn
&\times \left(-\tfrac{1}{64}(z-w)^{-\frac{9}{8}}(\bz-\bw)^{-\frac{9}{8}} \label{ae:o1o2}
+\tfrac{9}{128}(z-w)^{-\frac{5}{8}}(\bz-\bw)^{-\frac{5}{8}}\varepsilon_B(w,\bw)+\mathcal{O}(\p_{\{w,\bar{w}\}} \varepsilon_B)\right)\\
&+\left(-\tfrac{9}{64}(z-w)^{-\frac{1}{8}}(\bz-\bw)^{-\frac{17}{8}}
+\tfrac{15}{128}(z-w)^{\frac{3}{8}}(\bz-\bw)^{-\frac{13}{8}}\varepsilon_A(w,\bw)+\mathcal{O}(\p_{\{w,\bar{w}\}} \varepsilon_A)\right) \nn
&\times \left(-\tfrac{9}{64}(z-w)^{-\frac{1}{8}}(\bz-\bw)^{-\frac{17}{8}} \label{ae:o2o2}
+\tfrac{15}{128}(z-w)^{\frac{3}{8}}(\bz-\bw)^{-\frac{13}{8}}\varepsilon_B(w,\bw)+\mathcal{O}(\p_{\{w,\bar{w}\}} \varepsilon_B)\right) \,.
\end{align}
The calculation of the products ($\times$) and summing over Eqs.~\eqref{ae:o1o1}-\eqref{ae:o2o2} yields
\begin{align}
&\left[\mathcal{O}_{1}(z,\bar z)-\mathcal{O}_{2}(z,\bar z)\right]\left[\mathcal{O}_{1}(w,\bar w)-\mathcal{O}_{2}(w,\bar w)\right] \nn
=& \tfrac{81}{4096}\left((z-w)^{-\frac{17}{4}}(\bz-\bw)^{-\frac{1}{4}}+\text{H.c.}\right) 
- \tfrac{2}{4096}(z-w)^{-\frac{9}{4}}(\bz - \bw)^{-\frac{9}{4}} \nn
&- \left(\tfrac{135}{8192}\left((z-w)^{-\frac{15}{4}}(\bz-\bw)^{\frac{1}{4}}+\text{H.c.} \right) 
- \tfrac{9}{4096}(z-w)^{-\frac{7}{4}}(\bz-\bw)^{-\frac{7}{4}} \right) \left( \varepsilon_A+\varepsilon_B\right) \\
&+ \left(\tfrac{225}{16384} \left( (z-w)^{-\frac{13}{4}}(\bz-\bw)^{\frac{3}{4}}+\text{H.c.}\right) - \tfrac{81}{8192}(z-w)^{-\frac{5}{4}}(\bz-\bw)^{-\frac{5}{4}}\right) \varepsilon_A \varepsilon_B \\
&+ \mathcal{O}(\p_{\{w,\bar{w}\}} \varepsilon_I) \nonumber
\end{align}
which corresponds to the terms given in Eqs.~\eqref{e:alpha} and \eqref{e:gamma}.
\end{appendix}
\end{widetext}

\bibliographystyle{apsrev4-1}
\bibliography{Literature}

\end{document}